\documentclass[sigconf,nonacm]{acmart}

\usepackage{enumitem}
\usepackage{listings}

\setcopyright{none}
\begin{document}

\title{Whether Is Not When: Separating Occurrence from Timing in Polymarket Prices}

\author{Sizheng Fan}
\authornote{Corresponding author.}
\email{sizhengfan@uibe.edu.cn}
\affiliation{%
  \institution{University of International Business and Economics}
  \city{Beijing}
  \country{China}}

\author{Yinghe Sun}
\affiliation{%
  \department{CLNS}
  \institution{University of International Business and Economics}
  \city{Beijing}
  \country{China}}

\author{Rong Zhao}
\affiliation{%
  \department{Computer and Software}
  \institution{Shenzhen University of Information Technology}
  \city{Shenzhen}
  \country{China}}

\renewcommand{\shortauthors}{Fan et al.}

\begin{abstract}
Prices formed by Web users on Polymarket, a blockchain-based prediction market, are now cited in the news and in a U.S.\ Army intelligence bulletin as early warnings of geopolitical events. However, it remains unclear whether these prices can serve as such warnings. This cannot be judged from accuracy, because most markets resolve No and accuracy can remain high even when many of the events that did happen are missed. In this paper, we study this question from two aspects, using all 1.24 billion on-chain trades in 2.37 million Polymarket markets through September 2026. First, from 3,686 markets on geopolitical conflicts, we derive 208 unscheduled events that did happen and measure how far in advance prices move before the first Web report of each event. Second, we separate the information in prices about whether an event happens from the information about when it happens. To do so, we group 1,551 contracts into 338 deadline ladders and read each ladder as one forecast of the event date. We then split its log score, and the profit of 9.3 million trades by 491,000 accounts, exactly into these two parts. We find that one day before they were reported, prices correctly predicted only 56\% of the geopolitical events but all 51 scheduled central-bank decisions. Separating the two parts shows that only about one eighth of the information in prices concerns timing. A small group of large accounts earns timing profits, but beyond current prices, no group of accounts helps predict an event within days. A high price is therefore a sign that an event is likely rather than imminent. Near-term warnings read from the whole ladder are more often right than those read from a single contract. Platforms should report recall alongside accuracy, and show near-term probabilities read from the whole ladder.
\end{abstract}

\ccsdesc[500]{Applied computing~Economics}
\ccsdesc[300]{Information systems~Web mining}

\keywords{Web measurement, prediction markets, Polymarket, event timing, early warning, informed trading}

\maketitle

\section{Introduction}
\label{sec:intro}

On the Web, users now trade contracts on future events alongside stocks and cryptocurrencies.
These contracts are listed in prediction markets, where each contract pays out if a stated event happens by a deadline, so its price can be read as the probability of the event.
Polymarket is one of these markets.
Between November 2022 and September 2026, more than 3.1 million accounts traded contracts with a face value of about USD 110 billion in 1.24 billion trades on it, and every trade is recorded on a public blockchain.
At this scale, Polymarket prices have attracted wide attention.
In January 2026, Dow Jones announced that it would display Polymarket odds in its publications, including \textit{The Wall Street Journal} and \textit{Barron's}~\cite{dowjones2026polymarket}.
An article in the U.S.\ Army's \textit{Military Intelligence Professional Bulletin} also proposed such odds as leading indicators of security threats~\cite{ferris2025market}.
Through newspapers and intelligence bulletins, prices formed by Web users on one platform thus reach readers who do not trade on it, and are read as early warnings of events not yet reported.

However, it remains unclear whether these prices can serve as such warnings.
The evidence that Polymarket itself offers is accuracy.
In March 2025, it reported an accuracy of 90.4\% one month before resolution, counting a market as correct when its price lay on the side of 0.5 that matched the outcome~\cite{polymarket2025accuracy}.
Most markets, however, are on events that do not happen, and of the Yes/No markets settled on Polymarket by September 2026, 74\% resolved No.
A market on such an event is counted as correct as long as its price stays low, so a high accuracy says little about whether prices warned of the events that did happen.
Warnings are better judged by recall, the share of markets resolving Yes whose prices were at least 0.5 at a given time before the event was reported.

\begin{figure}[t]
  \centering
  \includegraphics[width=\columnwidth]{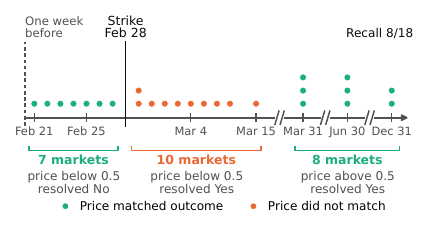}
  \Description{Schematic timeline of the 25 Polymarket markets on whether the U.S. would strike Iran by a given date that were still trading on February 21, 2026, one week before the strike, ordered by deadline. The 7 markets with deadlines from February 21 to 27 are green, priced below 0.5 and resolved No. The 10 markets with deadlines from February 28 to March 15 are orange, priced below 0.5 and resolved Yes. The 8 markets with deadlines on March 31, June 30 and December 31 are green, priced above 0.5 and resolved Yes. Recall is 8 of 18.}
  \caption{The 25 Polymarket markets on whether the U.S.\ would strike Iran by a given date that were still trading one week before the strike, ordered by deadline (not to scale). Green markets were priced on the side of 0.5 that matched the outcome, and orange markets were not.}
  \label{fig:iran}
\end{figure}

The U.S.\ strike on Iran on February 28, 2026 shows how low recall can be.
One week before the strike, 25 Polymarket markets were still asking whether the U.S.\ would strike Iran by a given date, and Figure~\ref{fig:iran} places them on a timeline by deadline.
The 7 markets with deadlines before the strike resolved No, and all of them were priced below 0.5.
The strike resolved the other 18 markets Yes.
At that time, however, only the 8 markets with deadlines at the end of March or later were priced above 0.5, so recall was only 8 of 18.
The February 28 contract was priced at 0.20, but this low price did not mean that traders saw a strike as unlikely, because the December 31 contract was priced at 0.75 at the same time.
Traders expected a strike within the year, but not by February 28.
The price of a single contract thus combines two questions, whether an event happens and when it happens.

Prior work on prediction markets has not split prices into these two questions.
In studies of accuracy and calibration, the price of each contract is scored against its own outcome~\cite{page2013calibrated,le2026decomposing,kagan2026calibration}.
Some of these studies group related contracts to estimate uncertainty~\cite{page2013calibrated,le2026decomposing}, but they do not read the contracts that ask one question under different deadlines, which we call a deadline ladder, as one forecast of the event date.
Studies of informed trading examine the behavior of wallets and the moves of prices before events~\cite{acdc2026insider,mitts2026iran}.
Mitts and Ofir~\cite{mitts2026iran}, for example, describe reports that in the hours before the strike on Iran, six new accounts earned about USD 1.2 million, mostly on the contract for a strike by February 28, a bet that turned on the date of the strike.
However, these studies do not split such profits into a part from knowing whether the event would happen and a part from knowing when.
Nechepurenko~\cite{nechepurenko2026foresightflow,nechepurenko2026population} measures the share of a contract's price move completed before the event, and for markets with deadlines compares it with a constant-hazard baseline for the time to the event.
This score is computed one contract at a time, so it does not read a deadline ladder as one forecast, and it could be obtained for only a small share of markets.
To the best of our knowledge, it therefore remains unknown, across a large set of events that did happen, how far in advance prices moved toward them and how much of what prices revealed concerned the date of the events rather than whether they would happen.

To address this gap, we study from two aspects whether Polymarket prices can serve as early warnings, using all 1.24 billion on-chain trades in its 2.37 million markets through September 2026.
First, we measure how far in advance prices move before events are reported.
From 3,686 markets on geopolitical conflicts, we derive 208 unscheduled events that did happen and date each one by the first Web report that meets the resolution rule of its market.
As a comparison, we use 51 central-bank rate decisions, whose dates traders know in advance.
Second, we separate the information in prices about whether an event happens from the information about when it happens.
We group 1,551 contracts into 338 deadline ladders and read the prices of each ladder on each day as one forecast of the event date (Figure~\ref{fig:framework}).
We then split the log score of this forecast exactly into these two parts, and split the profit of 9.3 million trades by 491,000 accounts in the same way.
We also compare warnings that an event will happen within seven days, read from the whole ladder, with warnings read from a single contract.

We find that one day before they were reported, prices correctly predicted only 56\% of the geopolitical events, but all 51 central-bank decisions.
Only about one eighth of the information in prices concerns when an event happens.
Prices thus carry far more information about whether an event happens than about when.
A small group of large accounts earns most of the profit from betting on the dates of events.
However, beyond current prices, their trades do not help predict whether an event will happen within days.
We also find that warnings of an event within seven days are more often right when read from the whole ladder rather than from a single contract.
Overall, a high price on Polymarket shows that an event is likely rather than imminent.
Our contributions are as follows.
\begin{itemize}[leftmargin=1.2em]
  \item We present a measurement framework that separates the two questions combined in the price of a prediction-market contract, whether an event happens and when it happens. The framework reads the contracts on one question as one forecast of the event date and splits both its log score and the profit of each trade exactly into the two parts. It trains no model and applies to any Web market that lists one question with several deadlines.
  \item We perform a large-scale measurement of prediction-market prices as early warnings, using all 1.24 billion on-chain trades on Polymarket and 208 geopolitical events dated by their first Web reports, and find that prices carry far more information about whether events happen than about when. Beyond current prices, no group of accounts helps predict whether an event will happen within days. We release the dated events, the deadline ladders and the inputs to all tests.
  \item We evaluate prices as early warnings for the platforms, newsrooms and intelligence analysts who read them, and offer recommendations to them. A high price should be read as a sign that an event is likely rather than imminent, and platforms should report recall alongside accuracy and show near-term probabilities read from the whole ladder rather than from a single contract.
\end{itemize}

\section{Background}
\label{sec:background}

In this section, we describe how Polymarket contracts are traded, how one question is listed under several deadlines, and how markets are resolved by an oracle.

\subsection{Contracts and Trading}
\label{sec:bg-contracts}

On Polymarket, each event page holds one or more binary markets, and each market has a Yes share and a No share~\cite{qin2026database,akey2026wins}.
One USDC, a stablecoin pegged to the U.S.\ dollar, can always be exchanged for a pair of Yes and No shares, and at resolution the winning share pays one USDC~\cite{akey2026wins,tsang2026anatomy}.
Under risk neutrality and ignoring the time value of money, the price of a Yes share is therefore the implied probability of the event~\cite{akey2026wins}.
In this paper, we also call a market a contract, and by its price we mean the price of its Yes share.

Orders are matched off chain in a central limit order book and settled on chain by a smart contract on the Polygon blockchain~\cite{qin2026database,akey2026wins}.
When an order to buy Yes is matched with an order to buy No, the smart contract creates a new pair of shares from their combined USDC, and a matched pair of orders to sell destroys a pair~\cite{akey2026wins,tsang2026anatomy}.
Every fill is recorded on the blockchain with its price, its amount and the addresses of the maker and the taker~\cite{qin2026database}.
We can therefore follow every trade of every account across all contracts on a question.

\subsection{Deadline Ladders}
\label{sec:bg-ladders}

Polymarket often lists one question with several deadlines as separate markets on the same event page.
The question \textit{US strikes Iran by\ldots?}, for example, was listed with 63 deadlines between January 11 and December 31, 2026.
These 63 markets form a deadline ladder.

On February 21, one week before the strike, the contract for February 28 traded at 0.20, the contract for March 31 at 0.63 and the contract for December 31 at 0.75 (Figure~\ref{fig:iran}).
A strike by February 28 is also a strike by December 31, so a contract with a later deadline pays out whenever one with an earlier deadline does, and prices should not fall as the deadline moves later.
The prices of a ladder on one day can therefore be read together as the probability that the event happens by each deadline.
The price of the latest deadline gives the probability that the event happens at all within the ladder, and the ratio of another price to it gives the probability that the event happens by that deadline if it happens.
On February 21, traders thus gave a strike a probability of 0.75 by December 31, and, if a strike came, a probability of about 0.27 (0.20/0.75) that it would come by February 28.
We formalize this reading in Section~\ref{sec:framework}.

\subsection{Resolution}
\label{sec:bg-resolution}

Each market states in text the rule that decides its outcome, including what counts as the event and, often, which sources settle it.
Once the outcome is known or the deadline passes, a proposer reports the outcome to UMA, an optimistic oracle on the blockchain, and any participant can dispute the proposal during a challenge window.
Disputes are decided by a vote of UMA token holders~\cite{akey2026wins,abid2026paradox}.
A market is therefore resolved some time after its event, and the delay depends on this process rather than on when the event became known.
For this reason, we date each event by the first Web report that meets the resolution rule of its market (Section~\ref{sec:data}) rather than by its resolution time.

\section{Data}
\label{sec:data}

In this section, we describe the trades and markets that we collected, how we dated events by Web reports, and how we grouped contracts into deadline ladders.

\subsection{Trades and Markets}
\label{sec:data-trades}

We collected all trades settled on chain by the two versions of the Polymarket exchange, from the first settlement on November 21, 2022 to September 25, 2026.
For the first version, we used the cleaned trade records of the Polymarket-v1 database~\cite{qin2026database}, which give the true direction of each trade, and we cleaned the trades of the second version in the same way.
From the Polymarket API, we also collected the question, resolution rule, deadline and outcome of each market.
In total, the data contain 1.24 billion trades in 2.37 million markets by 3.17 million accounts.
Following Akey et al.~\cite{akey2026wins}, we count each share at its face value of one dollar, which gives a trading volume of about USD 110 billion.
We take the price of a contract at time $t$ to be the price of its last trade before $t$, after removing wash trades (Section~\ref{sec:fw-wash}) and trades below one dollar, and we take the median when several trades share the last block.

\subsection{Dating Events by Web Reports}
\label{sec:data-events}

We built the events from the 3,686 markets on geopolitical conflicts.
We kept the markets that resolved Yes and whose outcome a single report can settle, such as a report that a strike took place.
Several of these markets often resolve Yes because of the same occurrence.
The strike on Iran on February 28, for example, settled all contracts on it with deadlines on or after that day, so we count it as one event.
We kept the onsets and turning points whose timing matters to the hour, and excluded events that were scheduled in advance or turned on statements, maps, elections or domestic politics, which left 256 events (Appendix~\ref{app:funnel}).

For each event, we define the time $\tau$ as the moment at which the first public report that meets the resolution rule of its market appeared on the Web.
To find this report, we ran an LLM agent with Web search twice for each event, each time independently, and asked it for the earliest qualifying report, its URL and the time shown on the page.
When the two searches disagreed, a third search received both answers and decided between them (Appendix~\ref{app:tau}).
We also checked by hand the times found for the first 200 events that we searched.
We dropped the events whose times still disagreed by more than six hours or conflicted with the market, and kept 208 events in 354 markets.
Deciding by $\tau$ the outcome of each deadline on the ladders of these events reproduces the resolution of the platform for 98\% of their markets.
As a comparison, we used 51 rate decisions of five central banks, including the Federal Reserve and the European Central Bank, between May 2024 and September 2026, each dated by its official announcement time.

\subsection{Deadline Ladders}
\label{sec:data-ladders}

We grouped markets into deadline ladders by removing the date phrase from each question, so that \textit{US strikes Iran by February 28, 2026?} and \textit{US strikes Iran by March 31, 2026?} fall into one ladder.
Each ladder includes its markets that resolved No, and we read each deadline from the question and the resolution rule.
Among the geopolitical markets that ask whether an event happens by a date, this gives 338 ladders with 1,551 contracts.
We call one ladder on one day a ladder-day.
Of these ladders, 282 have at least two deadlines with recent prices on some day, giving 12,573 ladder-days (Section~\ref{sec:fw-dist}).
Table~\ref{tab:sample} summarizes the samples used in each analysis.

\begin{table}[t]
  \caption{Samples used in each analysis.}
  \label{tab:sample}
  \small
  \begin{tabular}{@{}p{0.30\columnwidth}p{0.64\columnwidth}@{}}
    \toprule
    Analysis & Sample \\
    \midrule
    Whole platform & 1.24 billion trades, 2.37 million markets, 3.17 million accounts \\
    How far in advance & 208 geopolitical events in 354 markets; 51 central-bank rate decisions \\
    Whether versus when & 282 ladders, 12,573 ladder-days \\
    Accounts & 338 ladders, 1,551 contracts, 9.3 million trades, 491,000 accounts \\
    Warnings & 12,573 ladder-days, 494 followed by an event within seven days \\
    \bottomrule
  \end{tabular}
\end{table}

\subsection{Ethics}
\label{sec:data-ethics}

All data that we use are public.
Trades are recorded on a public blockchain, and the LLM agent read only pages that anyone can open.
We report accounts only in aggregate and do not name any address.
We release the dated events with the URLs of their reports but not the text of the news articles, together with the ladders and the inputs to all tests.

\section{Measurement Framework}
\label{sec:framework}

\begin{figure*}[t]
  \centering
  \includegraphics[width=\textwidth]{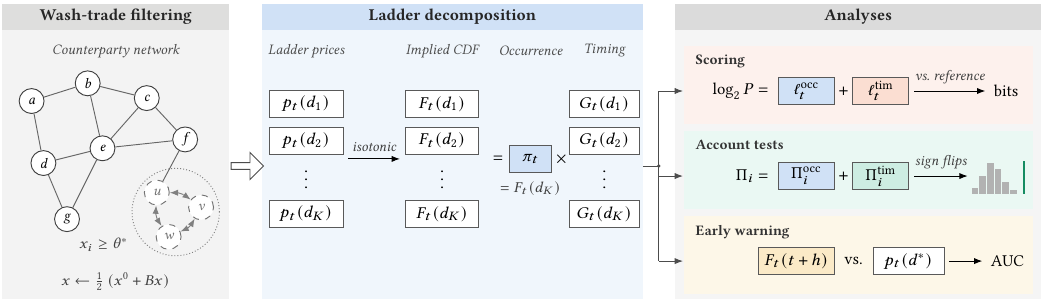}
  \Description{Overview diagram with three panels. Left: a counterparty network of accounts in which a cluster of wash-trading accounts is marked. Middle: the prices of a deadline ladder are mapped by isotonic regression to a distribution over the event date, written as an occurrence probability times a timing distribution. Right: three rows for scoring, account tests and early warning.}
  \caption{Overview of the measurement framework. After wash trades are removed, the prices of a deadline ladder on each day are fitted by isotonic regression to the event-date distribution $F_t$, which is split into the probability of occurrence $\pi_t = F_t(d_K)$ and the timing distribution $G_t$. The same split is applied to the log score and to the profit of each trade. As a warning, the probability of the event within $h$ days, $F_t(t+h)$, is compared with the price $p_t(d^*)$ of a single contract.}
  \label{fig:framework}
\end{figure*}

In this section, we read the prices of a deadline ladder on each day as one forecast of the event date, and use this forecast in three ways (Figure~\ref{fig:framework}).
We score how much information prices carry about whether and when the event happens, test which accounts bring the timing part into prices, and read near-term warnings from the whole ladder.
Before all three, we remove wash trades.

\subsection{Removing Wash Trades}
\label{sec:fw-wash}

Wash trades leave the positions of the traders unchanged, so they can move prices without carrying information.
We flag them with the network method of Sirolly et al.~\cite{sirolly2025wash} and keep its parameters.
Each account starts with a score $x^0$, the share of its traded shares that falls in markets where it closed its position, and its score is then updated on the network of counterparties until it converges,
\begin{equation}
x \leftarrow \tfrac{1}{2}\left(x^0 + Bx\right),
\label{eq:wash}
\end{equation}
where $B$ is the row-normalized matrix of shares traded between accounts.
In each market, we flag the accounts whose scores exceed a threshold chosen by the same method, and we treat trades between two flagged accounts, as well as self-trades, as wash trades.
Unlike Sirolly et al., we compute the scores only on the markets we study, where 13.8\% of traded shares are flagged.

\subsection{Event-Date Distribution of a Ladder}
\label{sec:fw-dist}

We form ladder-days at 00:00 UTC.
On day $t$, a contract enters its ladder if it opened before $t$, expires after $t$ and traded in the previous three days, and a ladder enters if at least two of its contracts with different deadlines enter.
When several contracts share a deadline, we keep the one with the largest volume.
Because $\tau$ is known only for the 208 dated events, we take the event of a ladder to become known at the price-implied time, when the price of a contract that resolved Yes first reaches 0.95 and never falls below 0.9 afterwards, and we use prices only before this time.
For the dated events, this time differs from $\tau$ by a median of two minutes.

\begin{definition}[Event-date distribution]
Let $d_1 < \dots < d_K$ be the deadlines of a ladder on day $t$, with prices $p_t(d_1), \dots, p_t(d_K)$.
The event-date distribution $F_t$ is the isotonic regression of these prices on the deadlines, so that $F_t(d_1) \le \dots \le F_t(d_K)$.
\end{definition}

We compute the isotonic regression with the algorithm of Ayer et al.~\cite{ayer1955empirical}.
It enforces the order described in Section~\ref{sec:bg-ladders}.
Prices rarely break this order, and on only 0.6\% of ladder-days is a later deadline cheaper than an earlier one by more than 0.02.
We then split $F_t$ into two parts,
\begin{equation}
F_t(d) = \pi_t G_t(d), \quad \pi_t = F_t(d_K), \quad G_t(d) = F_t(d)/\pi_t.
\label{eq:split}
\end{equation}
In other words, $\pi_t$ is the probability that the event happens by the latest deadline, and $G_t(d)$ is the probability that it happens by $d$ if it happens at all.
Split-population duration models separate whether and when in the same way~\cite{beger2016irregular}.
To read $F_t$ between two deadlines, we assume that an event that has not happened yet happens on each day with the same probability, so that the probability $S_t = 1 - F_t$ that the event has not happened is given by
\begin{equation}
S_t(s) = S_t(d_{k-1}) \left(\frac{S_t(d_k)}{S_t(d_{k-1})}\right)^{\frac{s - d_{k-1}}{d_k - d_{k-1}}}
\label{eq:interp}
\end{equation}
for $d_{k-1} \le s \le d_k$, with $d_0 = t$ and $S_t(t) = 1$.

\subsection{Splitting the Log Score}
\label{sec:fw-score}

Read this way, a ladder-day gives a full forecast of when the event happens.
The event either happens in an interval $(d_{j-1}, d_j]$, with probability $\pi_t (G_t(d_j) - G_t(d_{j-1}))$, or does not happen by $d_K$, with probability $1 - \pi_t$.
With $y = 1$ if the event happens by $d_K$, the log score $\log_2 P_t$ of the outcome splits exactly into two parts,
\begin{equation}
\begin{aligned}
\log_2 P_t &= W_t + T_t, \\
W_t &= y \log_2 \pi_t + (1 - y) \log_2 (1 - \pi_t), \\
T_t &= y \log_2 \bigl(G_t(d_j) - G_t(d_{j-1})\bigr).
\end{aligned}
\label{eq:logscore}
\end{equation}
Here $W_t$ scores whether the event happens and $T_t$ scores when, and $T_t$ is zero when the event does not happen.
We use the log score because it is strictly proper~\cite{gneiting2007strictly} and, unlike the squared error, splits exactly into these two parts.

We measure the information in each part as its average gain, in bits, over a reference forecast that lacks it.
We group ladders on the same conflict, such as the confrontation between the U.S.\ and Iran, into 45 storylines, fit each reference on other storylines and resample storylines for intervals.
For $W_t$, the reference is the base rate, the share of ladder-days with a latest deadline at a similar distance whose event happened.
For $T_t$, the main reference keeps $\pi_t$ but spreads it over the days until $d_K$ with the same daily probability,
\begin{equation}
F^{\mathrm{b}}_t(d) = 1 - (1 - \pi_t)^{(d - t)/(d_K - t)},
\label{eq:ref}
\end{equation}
so it knows whether the event happens as well as the market does, but nothing about its date.
A second reference replaces the constant daily probability with a Weibull distribution, so it also knows how fast such events usually happen, but still nothing about the event at hand.
We floor all probabilities at 0.005.
We score every ladder-day, including those of ladders whose event never happened.

\subsection{Splitting the Profit of Each Trade}
\label{sec:fw-profit}

The same split applies to the profit of each trade.
Consider a trade that changes the Yes holding of an account by $q$ shares on the contract with deadline $d_k$ at price $p$, and let $\pi \ge p$ be the price of the latest deadline at that moment and $G = p/\pi$.
With $y_k$ the outcome of the contract and $Y$ that of the latest deadline, the profit splits exactly as
\begin{equation}
q(y_k - p) = qG(Y - \pi) + qY(y_k - G).
\label{eq:profit}
\end{equation}
The first term is the profit from betting on whether the event happens, and the second is the profit from betting on when.
For each account that traded in at least five ladders, we test each part with sign randomization~\cite{gomezcram2026accuracy}, flipping the direction of all its trades in a ladder together 2,000 times.
We estimate the share of accounts with true skill after accounting for luck~\cite{barras2010false}, and check whether accounts selected as skilled on the earlier half of the ladders remain skilled on the later half.
Finally, we divide accounts into new accounts, accounts skilled at timing, the largest 1\% of accounts by volume and all other accounts.
For each group, we test out of sample whether its net buying on the nearest deadline over the previous 24 hours, beyond current prices, helps predict that the event happens within one or three days.

\subsection{Reading the Ladder as a Warning}
\label{sec:fw-warn}

On each ladder-day, a warning predicts whether the event happens within the next $h$ days, with $h = 7$ in the main text.
We compare four readings.
The first is the price of a single contract, the one with the first deadline on or after $t + h$, which is what a reader sees on the page of one market.
The second is $\pi_t$ alone.
The third is $F_t(t + h)$, read from the whole ladder by Eq.~\eqref{eq:interp} without training any model.
The fourth is a gradient-boosted model trained on features of the whole ladder in other storylines.
We evaluate the readings by their area under the ROC curve, average precision and precision at fixed recall, with intervals from resampling storylines 1,000 times.

\section{How Far in Advance Prices Move}
\label{sec:advance}

In this section, we measure how far in advance prices moved toward the 208 geopolitical events, and compare them with the 51 central-bank decisions.
For each event, we follow the contract with the largest volume among those that it resolved Yes, and we say that prices correctly predicted the event at a given time if the price of this contract was at least 0.5.

\begin{figure}[t]
  \centering
  \includegraphics[width=\columnwidth]{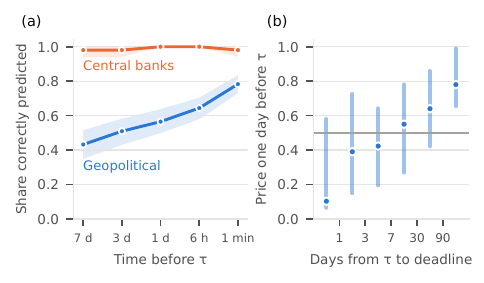}
  \Description{Two panels. Left: the share of correctly predicted events from seven days to one minute before the first report, near one for central-bank decisions and rising from about 0.43 to 0.78 for geopolitical events. Right: the median price one day before the first report rises from 0.10 to 0.78 as the deadline lies farther from the event.}
  \caption{Prices before the first report of each event. (a) Share of events whose contract was priced at 0.5 or higher, among the events whose contract was trading, with 95\% bootstrap intervals. (b) Median and interquartile range of the price one day before $\tau$ of the 312 contracts on the geopolitical events, grouped by the number of days from $\tau$ to the deadline.}
  \label{fig:advance}
\end{figure}

Figure~\ref{fig:advance}a shows how the share of correctly predicted events changed as the events approached.
One day before $\tau$, prices correctly predicted 109 of the 193 geopolitical events whose contracts were trading, or 56\%, but all 51 central-bank decisions.
Seven days before, the shares were 43\% and 98\%, and only 132 of the geopolitical events had a contract trading, because the contracts of the other 76 opened within a week of the event.
The share one day before was 56\% among these 132 events and 58\% among the 170 whose searches agreed within one hour (Appendix~\ref{app:robust}).
For the geopolitical events, the share rose to 64\% six hours before $\tau$ and to 78\% one minute before, so many prices moved toward the event only in its last hours.
These results indicate that prices anticipated the scheduled decisions long before their announcements, but only about half of the unscheduled events one day before these events were first reported on the Web.

How far in advance prices moved also differed across types of events.
One day before $\tau$, the median price was 0.97 for the 61 talks, meetings and calls and 0.76 for the 18 sanctions and diplomatic measures, but only 0.32 for the 49 military strikes and the 22 ceasefires.
In other words, prices warned of diplomatic events well before they were reported, but gave little warning of strikes and ceasefires.

To separate the type of event from the volume and the deadline of its contract, we regress the price one day before $\tau$ on the type of event, the logarithms of the volume before that day, of the age of the contract and of the number of days from $\tau$ to the deadline, and whether the event had a ladder, with intervals from resampling storylines.
Talks and calls were priced 0.37 higher than strikes (95\% interval 0.24 to 0.48), and a contract whose deadline fell 30 days after $\tau$ was priced about 0.26 higher than one whose deadline fell on the next day (0.07 to 0.37).
The intervals for volume, age and having a ladder all include zero.
We thus find no sign that contracts with more trading warned of their events earlier.

Figure~\ref{fig:advance}b shows the effect of the deadline directly, grouping the 312 contracts on these events by the number of days from $\tau$ to their deadlines.
One day before $\tau$, the median price was 0.10 for contracts whose deadline fell within one day after $\tau$, and rose to 0.78 for those whose deadline fell 90 days or more after $\tau$.
The same pattern holds within an event.
For the 72 events with at least two deadlines trading one day before $\tau$, the median price was 0.43 for the nearest deadline, a median of 1.8 days after $\tau$, and 0.72 for the farthest, a median of 36 days after $\tau$.
As in the strike on Iran (Section~\ref{sec:intro}), traders often expected these events to happen, but not by the nearest deadline.
At other times, near deadlines produced warnings that came too early.
Of the 180 contracts on the ladders of these events that resolved No and expired within 90 days before the event, 35 closed at 0.5 or higher on some day, in 21 of 57 events, although none was priced that high one day before its deadline.

In summary, one day before they were reported, prices correctly predicted only about half of the geopolitical events, and the price of each contract depended on how far its deadline lay from the event.
In the next section, we therefore separate how much of the information in prices concerns whether an event happens and how much concerns when it happens.

\section{Whether Versus When}
\label{sec:whether}

Section~\ref{sec:advance} showed that the price of a single contract depends on how far its deadline lies from the event.
In this section, we split the information in the prices of each ladder into whether and when the event happens, and then examine which accounts earn the profit from betting on when.

\subsection{Information About Whether and When}
\label{sec:whether-info}

\begin{figure}[t]
  \centering
  \includegraphics[width=\columnwidth]{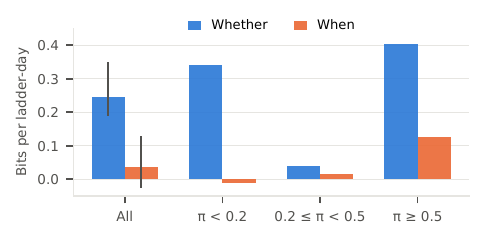}
  \Description{Grouped bars of bits per ladder-day for whether and when. Over all ladder-days, whether gives about 0.25 bits and when about 0.04 bits. Grouped by the probability of occurrence, when is negative below 0.2, near zero between 0.2 and 0.5, and about 0.13 bits at 0.5 or above.}
  \caption{Information in prices about whether and when the event happens, in bits per ladder-day beyond the reference forecasts, over all ladder-days and grouped by the probability of occurrence $\pi_t$. Lines show 95\% intervals over all ladder-days from resampling storylines.}
  \label{fig:bits}
\end{figure}

Figure~\ref{fig:bits} shows how much information prices carried about each part.
Over the 12,573 ladder-days, prices gave 0.25 bits per ladder-day about whether the event would happen beyond the base rate, which lowers the log loss of this part by 28\% (storyline interval 23\% to 37\%).
About when the event would happen, prices gave only 0.04 bits per ladder-day beyond the constant daily reference, which lowers the log loss of this part by 9\% (\textminus7\% to 26\%).
Even on the 4,192 ladder-days whose event happened, prices gave only 0.11 bits about when.
These findings indicate that only 13\% of the information in prices, about one eighth, concerned the date of the event.
Against the Weibull reference, prices lowered the log loss of the timing part by 18\% (8\% to 26\%), so learning how fast events usually happen placed the dates even worse than a constant daily probability.

The timing information came almost entirely from days on which prices already gave the event a probability of at least one half.
On the 3,570 ladder-days with $\pi_t \ge 0.5$, prices gave 0.13 bits about when, which accounts for 96\% of all timing bits.
On the 4,335 ladder-days with $\pi_t < 0.2$, the timing bits were negative, so when an event still looked unlikely, prices placed its date worse than a constant daily probability.
In other words, the date of an event became readable from prices only once prices had shown that the event was likely, so the timing part added little warning beyond the whether part.

The strike on Iran shows how prices misjudged a date in both directions.
On January 14, the contract for a strike by January 31 reached 0.78, and it later resolved No.
On January 30, the contract for February 28 traded near 0.6, but one day before the strike it had fallen to 0.07.
At that time, the 16 contracts that later resolved Yes but were priced below 0.5 all had deadlines within 16 days of the strike, and the 3 priced above 0.5 had deadlines a month or more after the strike.

Across events, the dates read from prices were also far from the actual dates.
Of the events in Section~\ref{sec:advance}, 50 had a ladder that asked whether the event would happen by each of at least two deadlines trading one day before $\tau$, and for these we read the date implied by prices as the first deadline priced at 0.5 or higher.
For 42 of them, this date fell a median of 18 days after the event happened and within three days of it for only 11.
For the other 8, no deadline was priced at 0.5 or higher.

\subsection{Accounts Betting on When}
\label{sec:whether-accounts}

We next examine who earns the profit from betting on when.
After we split the profit of the 9.3 million trades by Eq.~\eqref{eq:profit} and sum it in absolute value over accounts and ladders, the timing part amounted to USD 836 million, more than the USD 662 million of the whether part.
Although prices carried little information about the date, much money changed hands over it.

\begin{table}[t]
  \caption{Tests of whether accounts earn the two parts of profit by skill. The share with true skill is estimated after accounting for luck, over significance levels from 0.05 to 0.2.}
  \label{tab:accounts}
  \small
  \begin{tabular}{@{}lrr@{}}
    \toprule
    & Whether & When \\
    \midrule
    \multicolumn{3}{@{}l}{\textit{Accounts in at least five ladders}} \\
    Accounts tested & 71,133 & 13,241 \\
    Significant at 5\% & 5.0\% & 7.8\% \\
    With true skill & 0.0\% & 1.4--3.8\% \\
    \multicolumn{3}{@{}l}{\textit{Accounts in at least ten ladders}} \\
    Accounts tested & 20,069 & 2,651 \\
    Significant at 5\% & 6.3\% & 12.7\% \\
    With true skill & 1.6--1.7\% & 7.0--10.7\% \\
    \multicolumn{3}{@{}l}{\textit{Later half, accounts selected on the earlier half}} \\
    Still significant, selected & 12.7\% & 22.5\% \\
    Still significant, others & 6.4\% & 9.0\% \\
    \bottomrule
  \end{tabular}
\end{table}

Table~\ref{tab:accounts} tests whether some accounts earned these profits by skill rather than by luck.
Among the 13,241 accounts with timing profit in at least five ladders, 7.8\% earned a timing profit significant at the 5\% level, and we estimate that 1.4\% to 3.8\% have true skill in timing.
For the whether part, we estimate that almost no account has true skill.
Skill in timing is more common among accounts that traded in at least ten ladders, at 7.0\% to 10.7\%, and it persists, since 22.5\% of the accounts selected on the earlier half of the ladders remained significant on the later half, against 9.0\% of other accounts.

These accounts are few and large.
The 337 accounts with significant timing profit among those that traded in at least ten ladders had a median volume of USD 120,000, against USD 30,000 for other accounts, and they posted the order rather than taking it in 62\% of their volume, against 36\%.
The profit is also concentrated, and the top 1\% of the accounts with positive timing profit took 80\% of it.

However, the trades of these accounts did not reveal the date of an event before prices did.
We predicted whether the event happens within three days from current prices alone, and from current prices together with the net buying of each group of accounts over the previous 24 hours.
Out of sample, the area under the ROC curve was 0.935 with prices alone and 0.923 with net buying added.
Net buying by new and large accounts was slightly higher before events, but adding it never raised the area by more than 0.001, for one or three days and with or without the skilled accounts.
In other words, beyond current prices, no group of accounts helped predict whether an event would happen within days.

In summary, only about one eighth of the information in prices concerned when an event would happen, it appeared mostly once the event already looked likely, and one day before the events, the dates read from prices still fell weeks after them.
A small group of large accounts earned timing profits, but their trades added nothing to current prices.
In the next section, we examine how readers can still read near-term warnings from the whole ladder.

\section{Prices as Warnings}
\label{sec:warnings}

Section~\ref{sec:whether} showed that prices say little about when an event happens until the event already looks likely.
In this section, we examine how well prices can still warn of an event within seven days, and compare the four readings of Section~\ref{sec:fw-warn} over the 12,573 ladder-days, of which 494 (3.9\%) were followed by such an event.

\begin{figure}[t]
  \centering
  \includegraphics[width=\columnwidth]{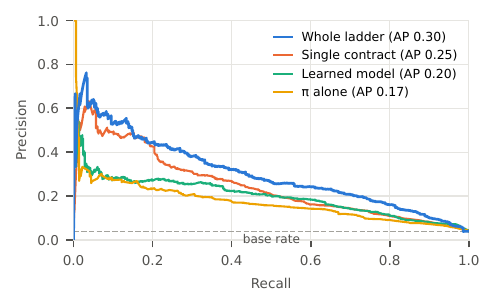}
  \Description{Precision-recall curves of four readings. The whole ladder lies above the single contract at most recall levels, and the learned model and the probability of occurrence alone lie below both. Average precision is 0.30 for the whole ladder, 0.25 for the single contract, 0.20 for the learned model and 0.17 for the probability of occurrence alone.}
  \caption{Precision and recall of four readings as warnings of an event within the next seven days, over 12,573 ladder-days. AP is the average precision.}
  \label{fig:warn}
\end{figure}

Figure~\ref{fig:warn} shows the precision and recall of the four readings.
The whole ladder gave the best warnings, with an area under the ROC curve of 0.88 and an average precision of 0.30, against 0.85 and 0.25 for the single contract.
The storyline intervals of these differences are 0.01 to 0.05 for the area and 0.01 to 0.11 for average precision.
When each reading caught half of the events, 26.5\% of the warnings read from the whole ladder were followed by the event, against 21.1\% of those read from the single contract.
The probability of occurrence $\pi_t$ alone gave the worst warnings, with an average precision of 0.17, because it says whether an event happens but not when.
The learned model did not improve on the whole ladder either, and its average precision was lower by 0.04 to 0.16.
The four readings rank in the same order for warnings within three days (Appendix~\ref{app:robust}).

The two main readings also differ for readers who treat a price of at least 0.5 as a warning.
The single contract with the first deadline on or after $t + 7$ also counts the chance that the event happens between $t + 7$ and its deadline, so as a warning of an event within seven days, its price is too high.
Read this way, the single contract warned of 43.5\% of the events within seven days, but only 24.1\% of its warnings were followed by the event.
The whole ladder warned of fewer events, 18.8\%, but 44.5\% of its warnings were right, and its Brier score was 0.033, against 0.062 for the single contract.

In summary, reading the whole ladder gives better near-term warnings than reading a single contract, and learning from other ladders adds nothing to this reading.
Even so, when the whole ladder caught half of the events within seven days, only about one in four of its warnings was right.

\section{Discussion}
\label{sec:discussion}

In this section, we discuss what our findings imply for the readers and platforms that treat prediction-market prices as warnings, and then turn to the limitations of our study.

\subsection{Implications}
\label{sec:disc-implications}

Polymarket prices are now cited in the news and proposed as indicators in intelligence work (Section~\ref{sec:intro}).
One day before the geopolitical events were reported, however, prices correctly predicted only 56\% of them, and only about one eighth of the information in prices concerned their dates (Sections~\ref{sec:advance} and~\ref{sec:whether}).
A high price should therefore be read as a sign that an event is likely rather than imminent.
Readers who need to know whether an event is near should read the whole ladder, since it gave better near-term warnings than any single price (Section~\ref{sec:warnings}).
Even then, when the whole ladder caught half of the events within seven days, three in four of its warnings were not followed by the event.

Polymarket reports the accuracy of its prices (Section~\ref{sec:intro}).
Because 74\% of its Yes/No markets resolved No, this accuracy says little about the events that did happen.
Platforms should therefore report recall alongside accuracy, such as the share of the events that did happen whose prices were at least 0.5 one day and one week before they were reported, which was 56\% and 43\% for the geopolitical events but 100\% and 98\% for the central-bank decisions.
Platforms already list the deadlines of one question on the same event page, so they can also show the probability that the event happens within the next week, read from the whole ladder by Eq.~\eqref{eq:interp}.
This reading trains no model and uses only the prices on the page.

The same reading applies to other uses of these prices on the Web.
Benchmarks of language-model forecasters evaluate forecasts or simulated trades on one Polymarket contract at a time~\cite{halawi2024approaching,cheng2026polybench,ye2026hindcast}.
When several contracts ask one question under different deadlines, evaluating them one at a time mixes whether and when, and the split of Section~\ref{sec:framework} can score the two separately.

Accounts that bet on the dates of strikes on Iran hours before them have raised concern about informed trading on Polymarket~\cite{mitts2026iran,acdc2026insider}.
We find that timing profits are concentrated in a small group of large accounts, but that across all ladders, no group of accounts helped predict whether an event would happen within days beyond current prices (Section~\ref{sec:whether-accounts}).
Timing profits are thus not by themselves a sign that traders knew the dates in advance.

\subsection{Limitations}
\label{sec:disc-limitations}

The primary limitation of our study is the timing of events.
We date each event by the first Web report found by an LLM agent, and we keep only the events whose searches agree within six hours, so our dates are precise to hours rather than minutes.
The rise in the share of correctly predicted events in the last six hours before $\tau$ (Section~\ref{sec:advance}) may thus partly reflect earlier reports that the searches missed.
Second, some of our samples are small.
For 76 of the 208 events, the contract with the largest volume opened within a week of the event, so we cannot measure warnings before it opened, and the 338 ladders come from only 45 storylines, so intervals from resampling storylines are wide, and the interval for the timing gain over the constant daily reference includes zero.
Third, one person can trade from several accounts, so the concentration of timing profit across accounts understates its concentration across people.
Finally, our study is descriptive.
We measure what prices revealed and when, but not why traders priced events as they did, and we study geopolitical events on one platform, leaving other kinds of events and other platforms to future work.

\section{Related Work}
\label{sec:related}

This section relates our work to prior research on 1) prediction-market prices as forecasts, 2) traders and information on Polymarket, and 3) forecasts of when conflicts begin.

\paragraph{Prediction-market prices as forecasts.}
Prediction markets aggregate dispersed information into prices~\cite{wolfers2004prediction,arrow2008promise}, and whether a price equals the average belief of traders depends on their budgets and preferences~\cite{manski2006interpreting,wolfers2006interpreting}.
The accuracy of these prices has been evaluated on election markets~\cite{berg2008results} and, more recently, on large samples of contracts~\cite{page2013calibrated,le2026decomposing,kagan2026calibration}.
Polymarket prices also circulate beyond the market.
Ibrahim and Zaki~\cite{ibrahim2026dislocations} find that a market is cited more often in the news in the days after a large price move, and benchmarks of language models evaluate their forecasts against the outcomes and prices of forecasting platforms and prediction markets such as Polymarket, or by the returns of simulated trades~\cite{halawi2024approaching,cheng2026polybench,ye2026hindcast}.
In these studies, each contract is scored or traded on its own, so the contracts that ask one question under different deadlines are not read together as a forecast of when the event happens.

\paragraph{Traders and information on Polymarket.}
Because every Polymarket trade is recorded on a public blockchain, it has become possible to ask where the information in its prices comes from.
Qin and Yang~\cite{qin2026database} release the complete trade archive of its first exchange, and such records have been used to correct its volume for minted and burned shares~\cite{tsang2026anatomy}, to show that trade directions inferred from the public order book are often wrong~\cite{dubach2026anatomy}, and to classify nearly 25\% of its historical volume as wash trading~\cite{sirolly2025wash}.
Profits on Polymarket are concentrated, as the top 1\% of users with positive profits capture 76.5\% of them~\cite{akey2026wins}, and a small minority of skilled accounts drives the accuracy of its prices~\cite{gomezcram2026accuracy}.
Kendler-Kretsch and Szakonyi~\cite{acdc2026insider} flag wallets and sudden price accelerations before news, Mitts and Ofir~\cite{mitts2026iran} flag accounts that traded before news, and Nechepurenko~\cite{nechepurenko2026foresightflow,nechepurenko2026population} measures how much of a contract's price move was completed before an event, with a constant-hazard baseline for markets with deadlines.
In these studies, profits and price moves are not split into a part about whether an event would happen and a part about its date.

\paragraph{Forecasts of when conflicts begin.}
Outside prediction markets, the timing of conflict has long been seen as hard to forecast.
Gartzke~\cite{gartzke1999war} argues that private information and incentives to bluff make the onset of war unpredictable, and Chadefaux~\cite{chadefaux2017market} finds that bond markets tended to underestimate the risk of war before it broke out and reacted with surprise once it began.
The ViEWS early-warning system forecasts political violence each month for countries and subnational areas~\cite{hegre2019views}, Mueller and Rauh~\cite{mueller2018reading} forecast conflict onset in each country a year ahead from newspaper text, and Mueller and Rauh~\cite{mueller2022hard} find onset hardest to forecast where there has been no recent violence.
Beger et al.~\cite{beger2016irregular} separate the two questions with split-population duration models, in which some variables predict whether a country is at risk of an irregular leadership change and others predict when it occurs.
In prediction markets, Leigh et al.~\cite{leigh2003iraq} read TradeSports contracts on the removal of Saddam Hussein with several expiry dates as the risk of war by each date, and used them to estimate the economic consequences of war.
Whether and when an event happens have thus been separated in models of conflict data, but the prices of markets that trade on such events have not been split into the two.

\section{Conclusion}
\label{sec:conclusion}

This paper measures whether prediction-market prices can serve as early warnings, separating the information in prices about whether an event happens from the information about when it happens.
Using all 1.24 billion on-chain trades on Polymarket through September 2026, 208 geopolitical events dated by their first Web reports and 338 deadline ladders, we find that one day before they were reported, prices correctly predicted only 56\% of the geopolitical events but all 51 central-bank decisions.
Only about one eighth of the information in prices concerned when events happen, and it appeared mostly once the events already looked likely.
One day before the events were reported, the dates read from prices still fell a median of 18 days after them.
A small group of large accounts earned most of the timing profit, but beyond current prices, no group of accounts helped predict events within days.
Near-term warnings read from the whole ladder were more often right than those read from a single contract.
A high price on Polymarket is therefore a sign that an event is likely rather than imminent, and platforms should report recall alongside accuracy and show near-term probabilities read from the whole ladder.
We release the dated events, the ladders and the inputs to all tests.

\bibliographystyle{ACM-Reference-Format}
\bibliography{references}

\appendix

\section{Dating Events by Web Reports}
\label{app:tau}

\begin{figure*}[t]
\begin{lstlisting}[basicstyle=\ttfamily\scriptsize,breaklines=true,columns=fullflexible,frame=single,framesep=4pt,xleftmargin=4pt,xrightmargin=4pt]
You are a careful fact-checker. Use web search and web fetch to find WHEN the following real-world event first became public. Do not rely on memory: every time you report must come from a page you actually opened.

Event (description in Chinese): {event}
Approximate date (UTC): {date}
Prediction markets resolved by this event, with their rules (they define exactly what must have happened):
{markets}

Two different times matter, never mix them:
- published_utc: when a post / statement / broadcast / article became visible to the public;
- event_utc: when the thing itself happened, as described by a source (e.g. "the strikes began at 22:40 GMT" written in a later article is an event time, NOT a publication time).

Rules for judging:
- satisfies_rules = true only if that item itself reports the event HAS happened AND meets the specific market rule (e.g. "missiles launched" does not satisfy "missiles entered Israeli airspace"; "talks scheduled" does not satisfy "met in person"; "deal close" does not satisfy "agreement reached").
- Consider ANY party involved: leaders' posts (Truth Social, X, Telegram), governments, militaries, state media of either side, mediators, live TV, wire flashes (Reuters/AP/AFP/Yonhap/etc.), local media in any language.
- For every time, state timestamp_basis: "page shows time with timezone", "page shows time, timezone inferred (say how)", "metadata/JSON-LD", "archive capture", "URL", or "other". Convert to UTC carefully.
- If a wire or news report says someone "said", "announced", "posted" or "confirmed" it (e.g. "Israel said it launched an attack", "Trump announced on Truth Social"), that original statement is usually earlier than the report: find the original post / statement / broadcast itself and its own time (the official's X or Telegram account, the ministry or military spokesperson channel, the live TV moment), including sources in Hebrew, Arabic, Persian, Russian, Ukrainian, Korean, etc. List both.
- Screenshots and reposts of social posts often show the viewer's local time without a timezone: only use such a time if the timezone can be established; otherwise use a source with an explicit timezone.
- If a page was updated later, say whether the quoted text was in the first version.
- If you cannot find something, use null. Never guess.

{hint}
Reply with ONLY one JSON object, no other text:
{"items": [{"published_utc": "YYYY-MM-DD HH:MM", "shown_time": "...", "timezone": "...", "timestamp_basis": "...", "event_utc": "... or null", "who_or_outlet": "...", "source_type": "leader post|official statement|state media|TV|wire flash|news article|other", "language": "...", "what": "one sentence", "satisfies_rules": true/false, "url": "...", "quote": "exact sentence or timestamp line"}],
 "first_public": {"published_utc": ..., "who_or_outlet": ..., "source_type": ..., "url": ..., "quote": ..., "timestamp_basis": ...},
 "second_source": {"published_utc": ..., "who_or_outlet": ..., "url": ..., "quote": ...},
 "earlier_signals": [{"published_utc": ..., "what": "public report before first_public indicating it was happening or imminent, but not yet satisfying the rules", "url": ...}],
 "per_market": [{"m": market number, "first_qualifying_published_utc": ..., "note": ...}],
 "event_utc": "when it happened, if stated, else null",
 "confidence": "high|medium|low", "notes": "..."}
first_public must be the item with the earliest published_utc among items with satisfies_rules = true (for the least strict market).
\end{lstlisting}
\caption{Prompt for each search. For each event, \{event\}, \{date\} and \{markets\} are filled in, and \{hint\} holds the notes for a third search and is empty otherwise.}
\label{fig:prompt}
\end{figure*}

We ran the searches with Composer 2.5, the agent model of the Cursor editor, in an isolated container with Web access, using the prompt in Figure~\ref{fig:prompt}.
For each event, the prompt gave a short description of the event, its approximate date, and for each of its markets the question, what the resolution rule requires to happen, what the rule does not count, and the time window of the rule.
The agent returned each report that it opened with the time shown on the page, its time zone and the basis of this time, and named the earliest report that meets the rule as the first public report.
Each event was searched twice, each time independently, and the two searches gave times within five minutes of each other for 47\% of the first 200 events and for 50\% of the other 96.

A third search used the same prompt with notes on the earlier answers.
When the two searches disagreed, the notes gave both answers and their sources and asked the agent to decide which one, if either, is the first report that meets the rule, or to find an earlier one.
A third search was also run when a time failed a check against the market and, for the last 96 events, when prices moved sharply in the two hours before the time found.
The checks reject a time that is missing, that falls before the first trade of the market or before the start of the window in its rule, or that falls after the end of this window or after the resolution of the market.
We also checked by hand the times found for the first 200 events.
We dropped the events whose times still differed by more than six hours or failed one of these checks.

Deciding the outcome of each deadline by $\tau$ reproduces the resolution of the platform for 345 of the 351 markets on the ladders of the dated events, and for 4,563 of their 4,623 deadlines on ladder-days.
The price-implied time of Section~\ref{sec:fw-dist} differs from $\tau$ by a median of two minutes.
For 157 of the 353 markets, the price-implied time came before $\tau$, a median of three hours earlier (Section~\ref{sec:disc-limitations}).

\section{Building the Events}
\label{app:funnel}

Table~\ref{tab:funnel} lists the steps from the markets on geopolitical conflicts to the 208 dated events.
We first kept the markets that resolved Yes and whose outcome a single report can settle, and merged the markets settled by the same occurrence into one event.
We then kept the onsets and turning points whose timing matters to the hour and excluded six kinds of events, and searched for the first report of each remaining event as in Appendix~\ref{app:tau}.
Finally, we removed one event whose question asks whether something does not happen, so that a resolution of Yes means that the event did not occur.

\begin{table}[h]
  \caption{Steps from the markets on geopolitical conflicts to the dated events.}
  \label{tab:funnel}
  \small
  \begin{tabular}{@{}p{0.62\columnwidth}rr@{}}
    \toprule
    Step & Markets & Events \\
    \midrule
    Markets on geopolitical conflicts & 3,686 & \\
    Resolved Yes and settled by a single report & 751 & 509 \\
    Onsets and turning points timed to the hour & 473 & 296 \\
    \quad Domestic politics or courts of other countries & & $-16$ \\
    \quad Control shown on maps of the ISW & & $-7$ \\
    \quad Statements & & $-6$ \\
    \quad Not geopolitical & & $-4$ \\
    \quad Scheduled in advance & & $-4$ \\
    \quad Elections & & $-3$ \\
    Searched for the first report & 422 & 256 \\
    \quad Times disagree or fail a check & & $-47$ \\
    \quad Question asked in the negative & & $-1$ \\
    Dated events & 354 & 208 \\
    \bottomrule
  \end{tabular}
\end{table}

\section{Squared-Error Version of the Split}
\label{app:brier}

With the squared error, the split of Eq.~\eqref{eq:logscore} is not exact, so we score the two parts separately.
For whether, we compare the squared error of $\pi_t$ on the latest deadline with that of the base rate.
For when, we compare the squared error of the prices on the other deadlines with that of the reference in Eq.~\eqref{eq:ref}, and with that of a reference learned by gradient boosting on other storylines.
We report the share by which prices lower the squared error of each reference.

Prices lowered the squared error by 33\% for whether (storyline interval 27\% to 42\%) and by 11\% for when (\textminus4\% to 21\%), and by 12\% for when against the learned reference (9\% to 22\%).
As with the log score, the timing gain came from days with $\pi_t \ge 0.5$, on which it was 14\%, against 4\% for $0.2 \le \pi_t < 0.5$ and \textminus3\% for $\pi_t < 0.2$.
On the ladders of 70 dated events, deciding the outcome of each deadline by $\tau$ rather than by the price-implied time gives a timing gain of 23\% (event interval 15\% to 32\%), against 18\% with the price-implied time on the same ladders.

\section{Robustness}
\label{app:robust}

For Section~\ref{sec:advance}, we restricted the events to those whose searches agreed more closely.
Among the 170 events whose searches agreed within one hour, prices correctly predicted 48\% seven days before $\tau$ and 58\% one day before, against 43\% and 56\% for all events.
Among the 126 events whose searches agreed within six minutes, the shares were 47\% and 61\%, and the share rose from 67\% six hours before $\tau$ to 75\% one minute before.

For Section~\ref{sec:whether}, flooring probabilities at 0.001 or 0.01 instead of 0.005 leaves the information about whether unchanged at 0.25 bits per ladder-day and puts the share of timing in all information at 12\% or 13\%.
For Section~\ref{sec:warnings}, the four readings rank in the same order for warnings of an event within three days, with an average precision of 0.28 for the whole ladder, 0.22 for the single contract, 0.20 for the learned model and 0.13 for $\pi_t$ alone.

\end{document}